\documentclass[ALICE,manyauthors]{cernphprep}
\usepackage[comma,square,numbers,sort&compress]{natbib}
\usepackage{hyperref}
\usepackage{lineno}
\usepackage{xspace}
\usepackage[T1]{fontenc}
\usepackage{orcidlink}

\begin{document}
%

\newcommand{\pp}           {pp\xspace}
\newcommand{\ppbar}        {\mbox{$\mathrm {p\overline{p}}$}\xspace}
\newcommand{\XeXe}         {\mbox{Xe--Xe}\xspace}
\newcommand{\PbPb}         {\mbox{Pb--Pb}\xspace}
\newcommand{\pA}           {\mbox{pA}\xspace}
\newcommand{\pPb}          {\mbox{p--Pb}\xspace}
\newcommand{\AuAu}         {\mbox{Au--Au}\xspace}
\newcommand{\dAu}          {\mbox{d--Au}\xspace}

\newcommand{\s}            {\ensuremath{\sqrt{s}}\xspace}
\newcommand{\snn}          {\ensuremath{\sqrt{s_{\mathrm{NN}}}}\xspace}
\newcommand{\pt}           {\ensuremath{p_{\rm T}}\xspace}
\newcommand{\meanpt}       {$\langle p_{\mathrm{T}}\rangle$\xspace}
\newcommand{\ycms}         {\ensuremath{y_{\rm CMS}}\xspace}
\newcommand{\ylab}         {\ensuremath{y_{\rm lab}}\xspace}
\newcommand{\etarange}[1]  {\mbox{$\left | \eta \right |~<~#1$}}
\newcommand{\yrange}[1]    {\mbox{$\left | y \right |~<~#1$}}
\newcommand{\dndy}         {\ensuremath{\mathrm{d}N_\mathrm{ch}/\mathrm{d}y}\xspace}
\newcommand{\dndeta}       {\ensuremath{\mathrm{d}N_\mathrm{ch}/\mathrm{d}\eta}\xspace}
\newcommand{\avdndeta}     {\ensuremath{\langle\dndeta\rangle}\xspace}
\newcommand{\dNdy}         {\ensuremath{\mathrm{d}N_\mathrm{ch}/\mathrm{d}y}\xspace}
\newcommand{\Npart}        {\ensuremath{N_\mathrm{part}}\xspace}
\newcommand{\Ncoll}        {\ensuremath{N_\mathrm{coll}}\xspace}
\newcommand{\dEdx}         {\ensuremath{\textrm{d}E/\textrm{d}x}\xspace}
\newcommand{\RpPb}         {\ensuremath{R_{\rm pPb}}\xspace}

\newcommand{\nineH}        {$\sqrt{s}~=~0.9$~Te\kern-.1emV\xspace}
\newcommand{\seven}        {$\sqrt{s}~=~7$~Te\kern-.1emV\xspace}
\newcommand{\twoH}         {$\sqrt{s}~=~0.2$~Te\kern-.1emV\xspace}
\newcommand{\twosevensix}  {$\sqrt{s}~=~2.76$~Te\kern-.1emV\xspace}
\newcommand{\five}         {$\sqrt{s}~=~5.02$~Te\kern-.1emV\xspace}
\newcommand{\twosevensixnn}{$\sqrt{s_{\mathrm{NN}}}~=~2.76$~Te\kern-.1emV\xspace}
\newcommand{\fivenn}       {$\sqrt{s_{\mathrm{NN}}}~=~5.02$~Te\kern-.1emV\xspace}
\newcommand{\LT}           {L{\'e}vy-Tsallis\xspace}
\newcommand{\GeVc}         {Ge\kern-.1emV/$c$\xspace}
\newcommand{\MeVc}         {Me\kern-.1emV/$c$\xspace}
\newcommand{\TeV}          {Te\kern-.1emV\xspace}
\newcommand{\GeV}          {Ge\kern-.1emV\xspace}
\newcommand{\MeV}          {Me\kern-.1emV\xspace}
\newcommand{\GeVmass}      {Ge\kern-.2emV/$c^2$\xspace}
\newcommand{\MeVmass}      {Me\kern-.2emV/$c^2$\xspace}
\newcommand{\lumi}         {\ensuremath{\mathcal{L}}\xspace}

\newcommand{\ITS}          {\rm{ITS}\xspace}
\newcommand{\TOF}          {\rm{TOF}\xspace}
\newcommand{\ZDC}          {\rm{ZDC}\xspace}
\newcommand{\ZDCs}         {\rm{ZDCs}\xspace}
\newcommand{\ZNA}          {\rm{ZNA}\xspace}
\newcommand{\ZNC}          {\rm{ZNC}\xspace}
\newcommand{\SPD}          {\rm{SPD}\xspace}
\newcommand{\SDD}          {\rm{SDD}\xspace}
\newcommand{\SSD}          {\rm{SSD}\xspace}
\newcommand{\TPC}          {\rm{TPC}\xspace}
\newcommand{\TRD}          {\rm{TRD}\xspace}
\newcommand{\VZERO}        {\rm{V0}\xspace}
\newcommand{\VZEROA}       {\rm{V0A}\xspace}
\newcommand{\VZEROC}       {\rm{V0C}\xspace}
\newcommand{\Vdecay} 	   {\ensuremath{V^{0}}\xspace}

\newcommand{\hyp}         {${}^{3}_{\Lambda}\mathrm{H}\xspace}
\newcommand{\ee}           {\ensuremath{e^{+}e^{-}}} 
\newcommand{\pip}          {\ensuremath{\pi^{+}}\xspace}
\newcommand{\pim}          {\ensuremath{\pi^{-}}\xspace}
\newcommand{\kap}          {\ensuremath{\rm{K}^{+}}\xspace}
\newcommand{\kam}          {\ensuremath{\rm{K}^{-}}\xspace}
\newcommand{\pbar}         {\ensuremath{\rm\overline{p}}\xspace}
\newcommand{\kzero}        {\ensuremath{{\rm K}^{0}_{\rm{S}}}\xspace}
\newcommand{\lmb}          {\ensuremath{\Lambda}\xspace}
\newcommand{\almb}         {\ensuremath{\overline{\Lambda}}\xspace}
\newcommand{\Om}           {\ensuremath{\Omega^-}\xspace}
\newcommand{\Mo}           {\ensuremath{\overline{\Omega}^+}\xspace}
\newcommand{\X}            {\ensuremath{\Xi^-}\xspace}
\newcommand{\Ix}           {\ensuremath{\overline{\Xi}^+}\xspace}
\newcommand{\Xis}          {\ensuremath{\Xi^{\pm}}\xspace}
\newcommand{\Oms}          {\ensuremath{\Omega^{\pm}}\xspace}
\newcommand{\degree}       {\ensuremath{^{\rm o}}\xspace}

\begin{titlepage}
\PHyear{2024}           
\PHnumber{187}        
\PHdate{11 July}  

\title{Rapidity dependence of antideuteron coalescence in pp collisions at \mbox{$\mathbf{\sqrt{\textit s} =}$ 13 TeV} with ALICE}
\ShortTitle{Rapidity dependence of antideuteron coalescence in pp at \mbox{$\sqrt{s} =$ 13 TeV}}   

\Collaboration{ALICE Collaboration\thanks{See Appendix~\ref{app:collab} for the list of collaboration members}}
\ShortAuthor{ALICE Collaboration}

\begin{abstract}
The production yields of antideuterons and antiprotons are measured in pp collisions at a center-of-mass energy of $\sqrt{s}= $ 13 TeV, as a function of transverse momentum (\pt) and rapidity ($y$), for the first time rapidity-differentially up to \mbox{$|y|=$ 0.7}. The measured spectra are used to study the \pt and rapidity dependence of the coalescence parameter $B_2$, which quantifies the coalescence probability of antideuterons. The \pt and rapidity dependence of the obtained $B_2$ is extrapolated for \pt $>$ 1.7 GeV/$c$ and $|y|>0.7$ using the phenomenological antideuteron production model implemented in \textsc{Pythia} 8.3 as well as a baryon coalescence afterburner model based on \textsc{EPOS} 3. 
Such measurements are of interest to the astrophysics community, since they can be used for the calculation of the flux of antinuclei from cosmic rays, in combination with coalescence models.
\end{abstract}
\end{titlepage}

\setcounter{page}{2}

\section{Introduction}
\label{sec:introduction}

Antinuclei such as antideuterons and antihelium still elude detection in space and, if discovered, would likely open the door to the indirect detection of weakly interacting dark-matter (DM) candidates~\cite{Donato:1999gy, Aramaki:2015pii}. Other scenarios speculate about the existence of anticlouds or antistars as another possible explanation of antinuclei in space~\cite{Poulin:2018wzu}. Indeed, the possible presence of antinuclei in our Galaxy could be explained either by reactions of high-energy cosmic rays (CRs) with the interstellar medium (ISM) or by more exotic sources such as decays/annihilations of DM candidates. Proton--proton (pp) and proton--nucleus (p--A) collisions are notably interesting in such a context, since the collisions between CRs and the ISM are the most relevant sources for the formation of nuclei in the Galaxy, because both, CRs and the ISM, consist mostly of hydrogen \mbox{($\sim$ 90$\%$)} and helium ($\sim$ 9$\%$), and only in small percentage of heavier nuclei.
The observation of a significant antimatter excess with respect to the expected background of antimatter produced in ordinary cosmic-ray interactions would represent a signal for dark-matter annihilation in the galactic halo or for the existence of antimatter islands in our Universe~\cite{Blum:2017qnn, Poulin:2018wzu}.

Information on the production of antinuclei from accelerator experiments, particularly in small collision systems (i.e., pp and p--A collisions), is essential for the theoretical description of the background constituted by antinuclei from cosmic-ray collisions with ISM. 
Since the ISM is almost at rest and the CRs have kinetic energies peaked at around 300 GeV, the values of relevant center-of-mass energies per nucleon--nucleon collision for the antideuteron production range from \snn$\sim$17 GeV to several TeV~\cite{Serksnyte}. 

The goal of the existing experimental programs at different accelerator facilities is to obtain antinucleus production cross sections for pp and p--A reactions over a broad range of kinetic energies, to cover as much as possible the energy range spanned by the cosmic rays. To this end, several experiments have measured the production of (anti)nuclei in different collision systems and center-of-mass energies. These include the low collision energies of the AGS~\cite{ Bennett:1998be, Ahle:1999in, Armstrong:2000gz, Armstrong:2000gd}, SPS~\cite{ Ambrosini:1997bf}, and RHIC~\cite{Adler:2001prl,Adler:2004uy, Arsene:2010px,Agakishiev:2011ib,Adamczyk:2016gfs,Adam:2019wnb}, and the TeV scale energies of the Large Hadron Collider (LHC)~\cite{Adam:2015pna, 3HePbPb, Acharya:2017dmc, deuteronFlowALICE502TeV, Acharya:2017fvb, Acharya:2017bso, LHCb:2018ygc, deuteron_pp7TeV, Acharya:2019rys, Acharya:2019ttn, Acharya:2019xmu, Acharya:2020sfy, Acharya:2020lus, Acharya:2020ogl, ALICE:2021mfm, nuclei_pp_5TeV, hypertriton_pPb_ALICE, ALICE:2022boh, ALICE:2022weh, ALICE:2022ugx, ALICE:2022zuz}. Such measurements are crucial for correct interpretations of any future measurement in satellite and balloon-borne experiments, such as AMS-02~\cite{Kounine:2012ega}, GAPS~\cite{Hailey:2009fpa}, or BESS-Polar~\cite{Abe:2011nx}. 
An ultimate goal of the experimental programs at accelerators is to pin down the microscopic production process of antinuclei in hadronic collisions or DM decay. In fact, the production mechanism of light (anti)nuclei in high-energy hadronic collisions is still not clear and much debated in the scientific community. The plethora of experimental data collected in the last decade is typically described using two different phenomenological models: the statistical hadronization model (SHM) and the baryon coalescence approach. In the SHM~\cite{SHM1,SHM2,SHM3,SHM4,SHM5,SHM6,SHM7}, light (anti)nuclei, as well as other hadron species, are assumed to be emitted by a source in local thermal and hadrochemical equilibrium with their abundances being fixed at the moment of the chemical freeze-out of the system created in the collision, at a temperature of \mbox{$T_{\rm chem}\sim $ 156 MeV}~\cite{Andronic:2018qqt}. This model provides an excellent description of the measured hadron yields in central nucleus--nucleus collisions~\cite{SHM5}. However, SHM struggles to reproduce the evolution with the charged-particle multiplicity of the ratios between integrated yields of nuclei and those of protons, especially for nuclei with masses A$\geq$3~\cite{ALICE:2022weh,ALICE:2022boh,ALICE:2024say}. In the coalescence model~\cite{Coalescence1,Coalescence2,Coalescence3,Coalescence4,Coalescence5,Coalescence6,Coalescence7}, multi-baryon states are assumed to be formed by coalescence of baryons that are close in phase space at kinetic freeze-out, which occurs at a later time than the chemical freeze-out. In the simplest implementations of coalescence, only the momentum correlations are considered and the bound states are formed if the difference in momentum among the nucleons lies below a given threshold, namely the coalescence momentum $p_0$. 
In the state-of-the-art implementations of the coalescence approach, the quantum-mechanical properties of the constituent baryons and of the final bound states are taken into account and the coalescence probability is calculated from the overlap between the wave functions of individual (point-like) baryons and the Wigner density of the final-state cluster~\cite{Mahlein:2023fmx}. 

From the experimental point of view, the coalescence probability is related to the coalescence parameter $B_A$, which is obtained from the ratio of the invariant yield of the nuclei with mass number $A$ and that of the protons raised to the power of $A$, assuming protons and neutrons to have the same transverse momentum distributions as they are isospin partners, and $p_{\mathrm T}^{\mathrm p} = p_{\mathrm T}^{A}$/$A$

\begin{equation}
B_{A} = { \biggl( \dfrac{1}{2 \pi p^{A}_{\mathrm T}} \biggl( \dfrac{ \rm{d}^2\it{N}} {\rm{d}\it{y}\rm{d}\it{p}_{\mathrm T}} \biggr)_{A}  \biggr)}  \bigg/{  \biggl( \dfrac{1}{2 \pi p^{\rm p}_{\mathrm T}} \biggl( \dfrac{\rm{d}^2\it{N}}{\rm{d}\it{y}\rm{d}\it{p}_{\mathrm T}}  \biggr)_{\rm p}  \biggr)^A}   .
\label{eq:BA}
\end{equation}

The production of (anti)nuclei at the LHC was so far measured in the midrapidity region $\left|y\right|<0.5$, and then, combined with several coalescence models, employed to predict the flux of antinuclei from CR interactions at forward rapidity. However, the possible impact of a rapidity dependence of the production yield and of the coalescence probability of antinuclei was suggested in Ref.~\cite{Citron:2018lsq}. Previous measurements of the production of (anti)protons and (anti)deuterons in different rapidity intervals have been carried out in heavy-ion collisions at RHIC energies by the BRAHMS Collaboration~\cite{Arsene:2010px}. In this Letter we report the first measurements of the rapidity dependence of the production yield of $\overline{\rm p}$ and $\overline{\rm d}$ and of the coalescence parameter $B_2$ (obtained from Eq.~\ref{eq:BA} with $A=2$) carried out in pp collisions at LHC energies. 

The results presented in this Letter contribute to the understanding of the impact that the extrapolation of the production yield and coalescence probability at forward rapidity, constrained to experimental information, has on the flux of antinuclei from cosmic rays, which can be obtained using hadronic production mechanisms based on coalescence. By measuring the production of antideuterons and antiprotons as a function of rapidity in pp collisions at $\sqrt{s}=$13 TeV up to $\left|y\right|=0.7$, the extrapolation at forward rapidity and high \pt$/A$ of the coalescence parameter $B_2$ is performed using several coalescence models. 
Given the sensitivity of the cosmic-ray experiments discussed above, i.e., AMS-02, GAPS, and BESS-Polar, and the current experimental uncertainties on the production measurements, the direct impact of these results on cosmological DM research is related to a better understanding of the antideuteron background flux~\cite{Gomez-Coral:2018yuk}, which needs to be precisely modeled to interpret future measurements of antideuteron CR flux correctly.
       
\section{Experimental apparatus}
\label{sec:Experiment}
ALICE is one of the four large experiments at the LHC and is dedicated to the study of hadronic collisions at ultra-relativistic energies. A detailed description of the ALICE apparatus and its performance can be found in Refs.~\cite{ALICEexperiment, ALICEperformance}. In the following, only the sub-detector systems used for this analysis are described.

Trajectories of charged particles are reconstructed in the ALICE central barrel, which covers the pseudorapidity interval of $|\eta| < 0.9$, with the Inner Tracking System (ITS)~\cite{ALICEexperiment}, the Time Projection Chamber (TPC)~\cite{TPC}, and the Time-Of-Flight (TOF) detector~\cite{TOF}. These detectors are located inside a solenoidal magnet, which generates a highly homogeneous magnetic field of 0.5 T, parallel to the beam line. 

The ITS consists of six cylindrical layers of silicon detectors and is used for the determination of primary and secondary vertices, and for charged-particle tracking.
The TPC is a gas detector used for charged-particle track reconstruction and momentum determination, and particle identification via the measurement of the specific energy loss (d$E$/d$x$) of particles in the detector gas. The \dEdx resolution depends on the event multiplicity and is about 5\%--6.5$\%$ for minimum-ionizing particles crossing the full volume of the TPC~\cite{ALICEperformance}. 
The particle identification is extended at high \pt using the TOF detector, which is located at a radial distance of 3.7 m from the nominal interaction point. It measures the arrival time of particles relative to the event collision time provided by the TOF detector itself or by the T0 detectors. The T0 consists of two arrays of Cherenkov counters, T0A and T0C, located on opposite sides of the interaction point, covering the pseudorapidity regions $4.6 < \eta < 4.9$ and $-3.3 < \eta < -3.0$. A weighted average is performed when both T0 and TOF detectors have measured the start time~\cite{CollisionTime}. The TOF time resolution is 56 ps~\cite{TOF}.

Collision events are triggered by two plastic scintillator arrays, V0A and V0C~\cite{VZEROPerformance}, located at asymmetric positions, one on each side of the interaction point, covering the pseudorapidity regions \mbox{$-3.7 < \eta < -1.7$}  and \mbox{2.8 $< \eta <$ 5.1}. Each V0 array consists of four rings in the radial direction, with each ring comprising eight cells with the same azimuthal size. The \VZERO detector is used to define the minimum-bias (MB) trigger (requiring coincident signals in the \VZERO detectors to be synchronous with the bunch-crossing time defined by the LHC clock). The V0 is also used to reject background events like beam--gas interactions, collisions with de-bunched protons, or with mechanical structures of the beam line~\cite{VZEROPerformance}.
       
\section{Data analysis}
\label{sec:DataAnalysis}

The analyzed data sample was collected in 2016, 2017, and 2018 during the LHC pp run at \mbox{$\sqrt{s} = $ 13 TeV}. 
Events with multiple vertices identified from track segments in the two innermost layers of the ITS are tagged as pile-up and removed from the analysis~\cite{ALICEperformance}. 
In order to ensure full geometrical acceptance in the ITS for $|\eta|<$ 0.9 and reject background collisions, the coordinate of the primary vertex along the beam axis is required to be within 10 cm from the nominal interaction point. 
A total number of approximately 1.7 billion MB pp events were analyzed, with a total integrated luminosity of about 22 $\rm{nb}^{-1}$~\cite{ALICE-PUBLIC-2021-005}.

The reconstructed tracks are required to fulfill the same set of quality criteria of the corresponding analyses of antiprotons and antideuteron production at midrapidity~\cite{ALICE:2020nkc,Acharya:2020sfy}, with the only difference that in the case of the present analysis no selections in pseudorapidity are applied, besides the ones enforced by the detector acceptance ($|\eta|<$0.9 in the ALICE central barrel). 
The antiproton (antideuteron) identification is done at \pt$<$ 0.7 GeV/$c$ (\pt  $< 1.2\ \mathrm{GeV}/\textit{c}$) by requiring that its energy loss per unit of track length measured by the TPC is within 3$\sigma_{\mathrm dE/\mathrm dx}$ from the expected average for antiprotons (antideuterons), where $\sigma_{\mathrm dE/\mathrm dx}$ is the \dEdx resolution. For \pt$>$ 0.7 GeV/$c$ for antiprotons and \pt  $> 1.2\ \mathrm{GeV}/\textit{c}$ for antideuterons, the \dEdx signal of the TPC is complemented by the time-of-flight measured by the TOF detector. 
The antiproton (antideuteron) signal is extracted from a fit to the n($\sigma^{\rm TOF}) = ( t^{\rm TOF} - t_{\mathrm{exp}}^{\rm TOF})/\sigma_{t}$ distribution, where $t^{\rm TOF}$ is the measured time-of-flight, $t_{\mathrm{exp}}^{\rm TOF}$ its expected value for protons (deuterons), divided by the resolution on the time-of-flight measurement ($\sigma_{t}$). The fit function consists of a Gaussian with an exponential tail for the signal and the sum of two exponential functions for the background. 
The raw signal yield is extracted by integrating the signal function in the asymmetric interval [$-3\sigma^{\rm TOF} + \mu_{0}, 3.5\sigma^{\rm TOF} + \mu_{0}$], where $\mu_{0}$ is the mean of the Gaussian. 
The acceptance of the ITS, TPC, and TOF detectors ($|\eta|<$0.9) limits the rapidity coverage of the measurement. Hence, the analysis is performed in seven rapidity intervals, 0.2 units wide, from $-0.7$ to 0.7. At $\left|y\right|$ larger than 0.7, the acceptance of the TPC and TOF detectors allows for the reconstruction of deuterons only at high \pt (for $\left|y\right|= 0.8$, the minimum \pt of deuterons that can be reconstructed is $\sim3.5$ GeV/$c$). However, in the high \pt region (\pt $>3.5$ GeV/$c$) deuterons cannot be identified any longer because the background due to mismatched tracks in the TOF is dominant with respect to the actual signal. The amount of observed raw antiprotons is about 2 $\times 10^{5}$ in each of the first five rapidity classes ($|y|<0.1$ to $0.4<|y|<0.5$), and decreasing to 1.5 $\times 10^{5}$ and 1 $\times 10^{5}$ in the two highest rapidity classes ($0.5<|y|<0.6$ and $0.6<|y|<0.7$, respectively). For antideuterons, the amount of observed raw counts is about 4 $\times 10^{4}$ in each of the first four rapidity classes ($|y|<0.1$ to $0.3<|y|<0.4$), and decreasing to 2 $\times 10^{4}$, 1 $\times 10^{4}$ and 5 $\times 10^{3}$ in the higher rapidity classes ($0.4<|y|<0.5$, $0.5<|y|<0.6$ and $0.6<|y|<0.7$, respectively). In both cases, the lower amount of observed antinuclei at high rapidity is due to the smaller transverse momentum coverage and lower acceptance and reconstruction efficiency. 

The raw antiproton and antideuteron \pt spectra are corrected for the acceptance and reconstruction efficiency and, only for antiprotons, for the fraction of secondary antiprotons produced by feed-down of weak decays of $\overline{\Lambda}$ hyperons. Both corrections are calculated with Monte Carlo (MC) simulations in which antinuclei are embedded into pp collision events generated using \textsc{Pythia} 8.1 with the Monash 2013 tune~\cite{PYTHIA8Monasch2013}. Antinuclei are generated with uniform \pt and rapidity distributions in  \mbox{(0 $<$ \pt $<$ 10) $\mathrm{GeV}/\textit{c}$} and $-1 < y < 1$. 
The particle interactions in the experimental apparatus are simulated using \textsc{Geant 4} as transport package~\cite{GEANT4}. The acceptance $\times$ efficiency, in each \pt and rapidity interval, is calculated as the ratio of reconstructed and generated antiprotons and antinuclei in the simulation, where the same track selection and PID criteria as those used in the data are applied to the reconstructed sample in MC. The resulting acceptance $\times$ efficiency depends on \pt, rapidity, and PID technique (either TPC only at low \pt, or TPC and TOF for \pt$>$ 0.7 GeV/$c$ for antiprotons, and for \pt$>$ 1.2 GeV/$c$ for antideuterons). For the class $|y|<0.1$, it varies from $\sim$50\% at low \pt to about 25\% at high \pt, for antiprotons, and from $\sim$80\% at low \pt to about 40\% at high \pt, for antideuterons. For the class $0.6<|y|<0.7$, instead, the PID is done only with the TOF, and the acceptance $\times$ efficiency varies from $\sim$10\% at low \pt to about 30\% at high \pt, for antiprotons, and from $\sim$20\% at low \pt to about 45\% at high \pt, for antideuterons.

In order to limit the contamination from secondary antiprotons from feed-down of weak decays, the measured distance of closest approach to the primary vertex in the transverse plane ($\mathrm{DCA}_{xy}$) and along the beam direction ($\mathrm{DCA}_{z}$) is requested to be less than 0.1 cm and 1 cm, respectively. To remove any remaining contamination of antiprotons from weak decays, the secondary fraction is estimated in each \pt and rapidity interval by fitting the distributions of the $\mathrm{DCA}_{xy}$, using the template method~\cite{3HePbPb}. The $\mathrm{DCA}_{xy}$ template distributions of primary (produced in the collision at the primary vertex) and secondary antiprotons from feed-down are taken from MC simulations. The resulting primary fraction of antiprotons has a mild \pt dependence, ranging from $\sim$80\% at low \pt to $\sim$90\% at high \pt. 

The sources of systematic uncertainties on the transverse momentum distributions considered for this analysis are related to (i) track selection and particle identification, (ii) detector material budget, (iii) TPC--ITS and TPC--TOF track matching efficiencies, (iv) hadronic interaction of antiparticles with the detector material, and (v) signal loss (to account for antiparticles lost in rejected events). The uncertainties related to (i) are estimated for the present analysis and discussed below, while the other ones are inherited, as relative uncertainties, from previous similar analyses. The uncertainties due to (ii) and (v) are estimated as described in Ref.~\cite{Acharya:2020sfy}, and those due to (iii) and (iv) are assessed as illustrated in Ref.~\cite{ALICE:2021mfm}. 

In order to assess the systematic uncertainties related to the track selection, the analysis is repeated using 50 settings with different criteria. These criteria are defined by randomly sampling the analysis parameters from uniform probability distributions, such that their variations produce a raw yield variation of a maximum factor of two, in order to uniformly explore the full available phase-space. The efficiency-corrected yields are distributed according to a Gaussian and the standard deviation of this distribution is taken as systematic uncertainty. 
The uncertainties connected to the particle identification are given by two contributions. One is estimated by extending or narrowing the integration intervals by $\pm 1\sigma$ and it is determined by calculating the standard deviation of the distribution of the fully corrected yields, in each \pt and rapidity interval, similarly to what was done in Ref.~\cite{Acharya:2020sfy}. In addition, it is included the small contribution ($\sim1\%$) due to the difference between the yield extracted by the integral of the fit function which describes the signal and that of the histogram after the subtraction of the background.
In the case of antiprotons, also the contribution related to the secondary fraction of antiprotons from weak decays is included in (i), hence the uncertainty is larger for antiprotons than for antideuterons. For this uncertainty, the DCA selection criteria (both in the perpendicular plane DCA$_{xy}$, and along the beam axis DCA$_{z}$) are also changed 50 times, and each time the corresponding primary fraction is used for correction. 
The total systematic uncertainties are obtained by summing in quadrature all the individual contributions, which are summarized in Table~\ref{Table:systematic_uncertainties}. Within the studied rapidity range, the systematic uncertainties have been found to be independent of the analyzed rapidity interval.

\begin{table}[!!htb]
\begin{center}
\centering
\renewcommand{\arraystretch}{1.4}
\caption{Summary of the systematic uncertainties on the transverse momentum distributions. The values reported in the table correspond to the uncertainties of the rapidity class $|y|<0.1$. For such class, low \pt corresponds to 0.375 GeV/$c$ for antiprotons and to 0.75 GeV/$c$ for antideuterons, whereas high \pt corresponds to 1.95 GeV/$c$ for antiprotons and to 3.2 GeV/$c$ for antideuterons. }
\begin{tabular}{lclcl}
\hline
Source of uncertainty     &  $\overline{\rm p}$ low \pt    &   $\overline{\rm p}$ high \pt &   $\overline{\rm d}$ low \pt    &   $\overline{\rm d}$ high \pt \\
\hline
Tracking and PID           &  4.5$\%$    &     6$\%$    &  1.5$\%$    &     3$\%$     \\
Material budget           &$<$1$\%$   &  $<$1$\%$    &$<$1$\%$   &   $<$1$\%$     \\
Matching efficiencies     &  3$\%$    &     3$\%$    &  1$\%$    &     2.5$\%$   \\
Hadronic interaction      &  2$\%$    &     2$\%$    &  3$\%$    &     4$\%$     \\
Signal loss               &  --    &     --    &  1$\%$    &     $<$1$\%$     \\
\hline
Total                     &  6$\%$    &     7$\%$    &  4$\%$    &     6$\%$     \\
\hline
\end{tabular}
\label{Table:systematic_uncertainties}
\end{center}
\end{table}

\section{Results}
\label{sec:Results}

\begin{figure}[!hbt]
\centering
\begin{minipage}[b]{0.49\textwidth}
	\includegraphics[width=\textwidth]{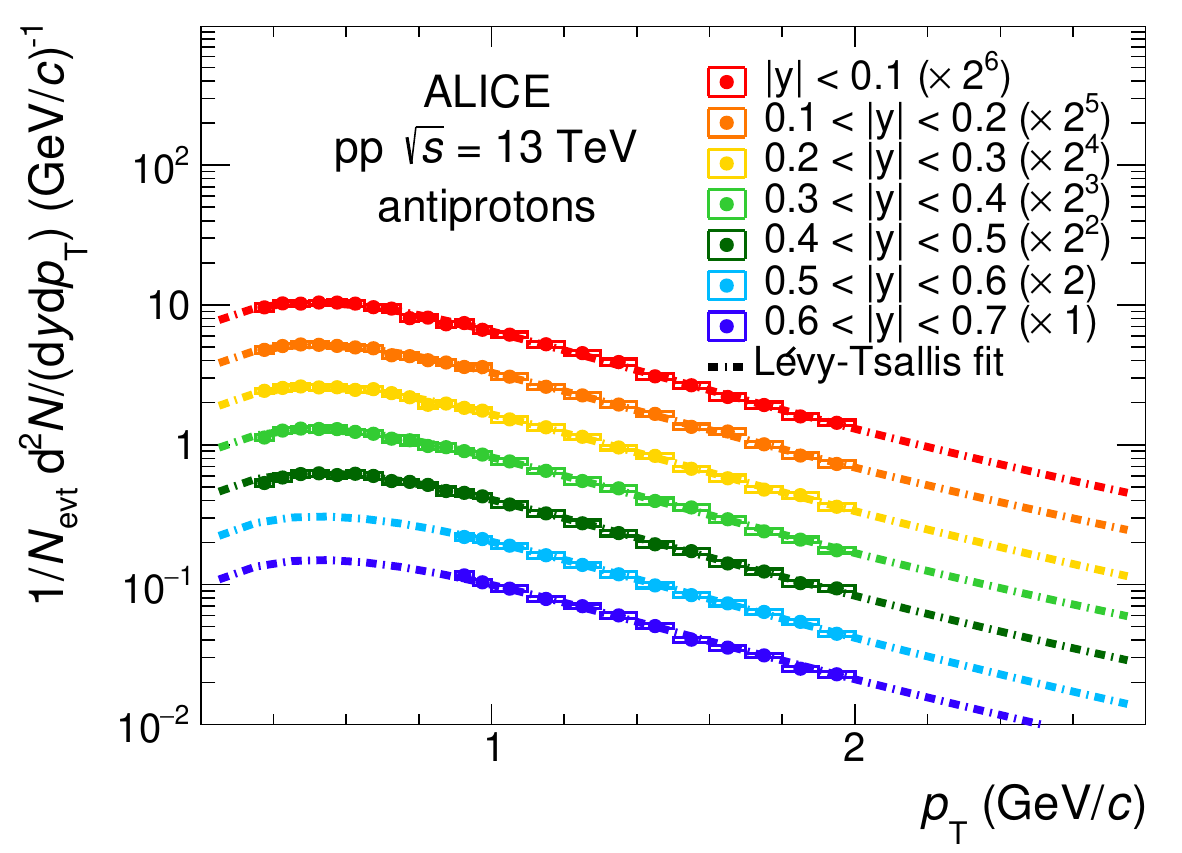}
\end{minipage}
  \hfill
\begin{minipage}[b]{0.49\textwidth}
    \includegraphics[width=\textwidth]{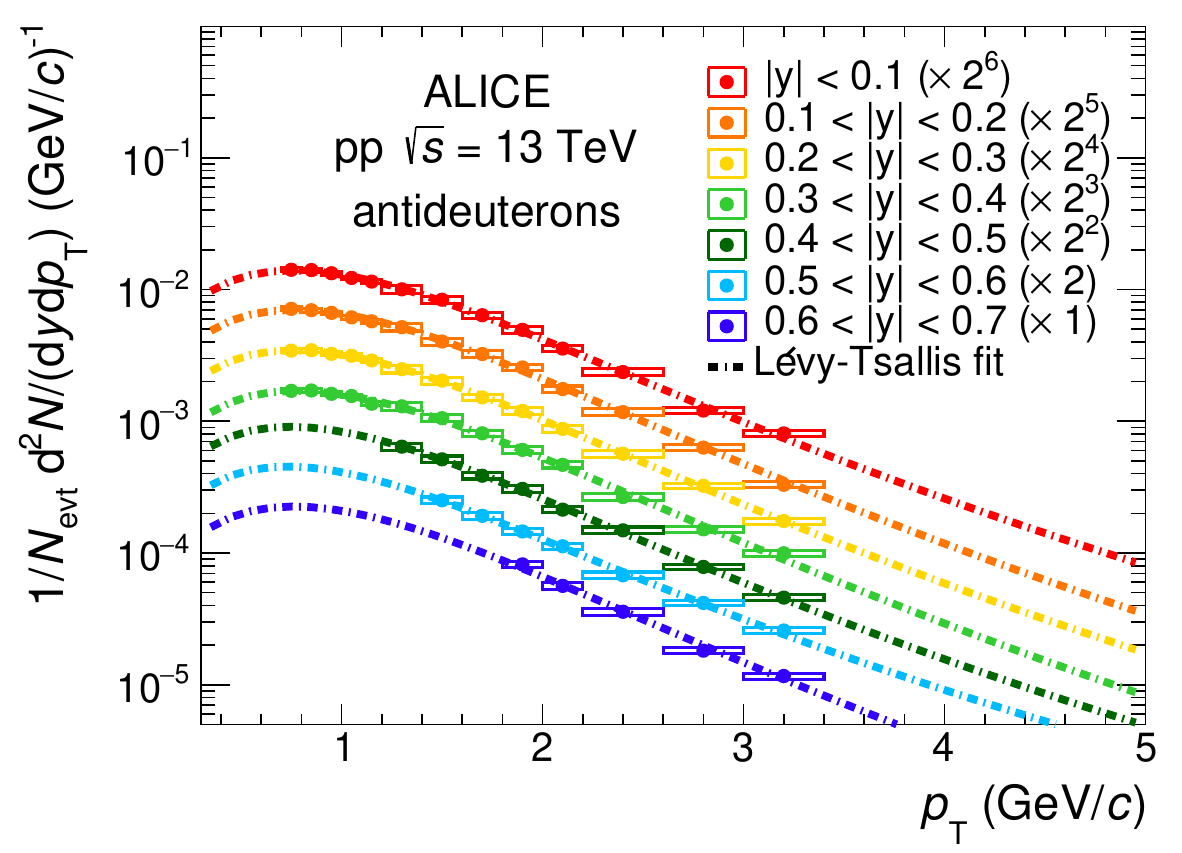}
\end{minipage}   
\caption{Antiproton (left panel) and antideuteron (right panel) \pt-differential yields for different rapidity intervals, in MB events. Statistical and systematic uncertainties are represented by vertical bars and boxes, respectively. The statistical uncertainties are smaller than the size of the markers in the reported scale and, hence, not visible. The dash-dotted lines represent the fit of the spectra executed with a L\'evy--Tsallis function. }
\label{fig:pt_spectra}
\end{figure}

\begin{figure}[!hbt]
\centering
\begin{minipage}[b]{0.49\textwidth}
	\includegraphics[width=\textwidth]{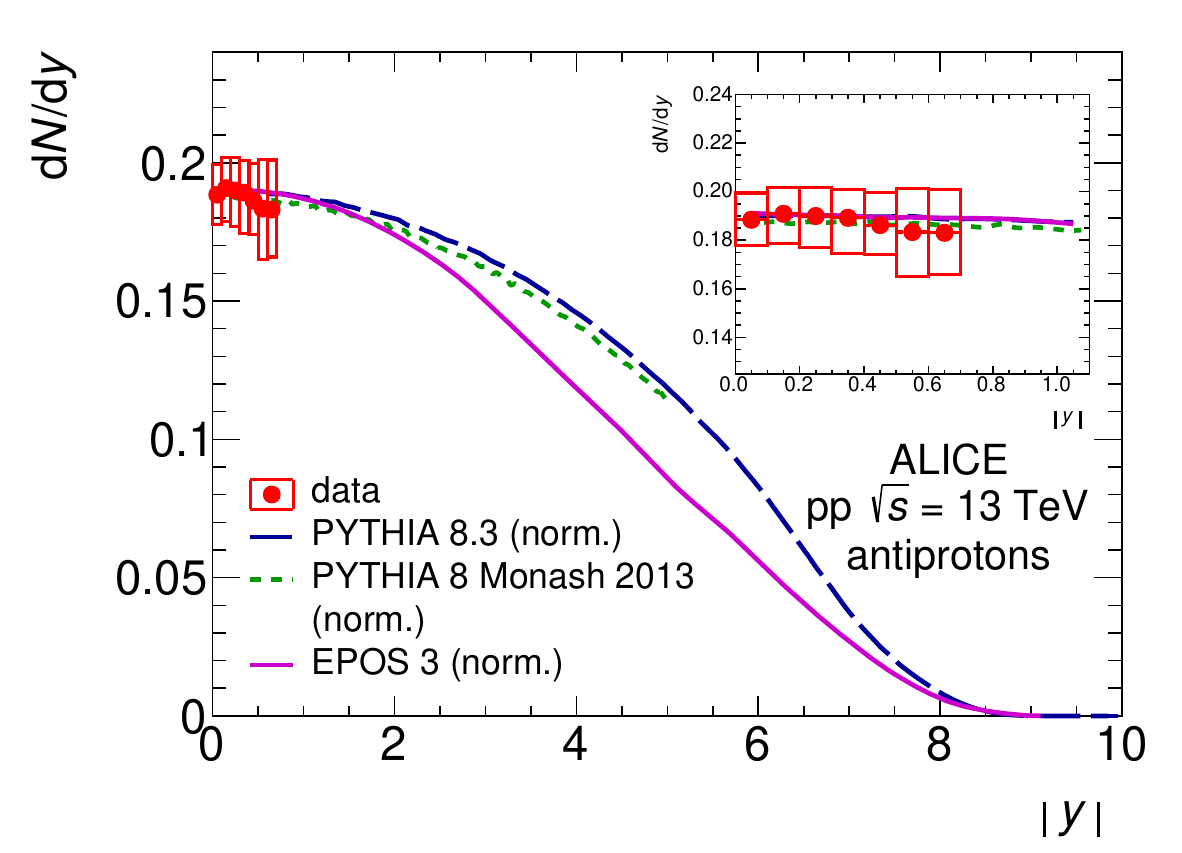}
\end{minipage}
\begin{minipage}[b]{0.49\textwidth}
	\includegraphics[width=\textwidth]{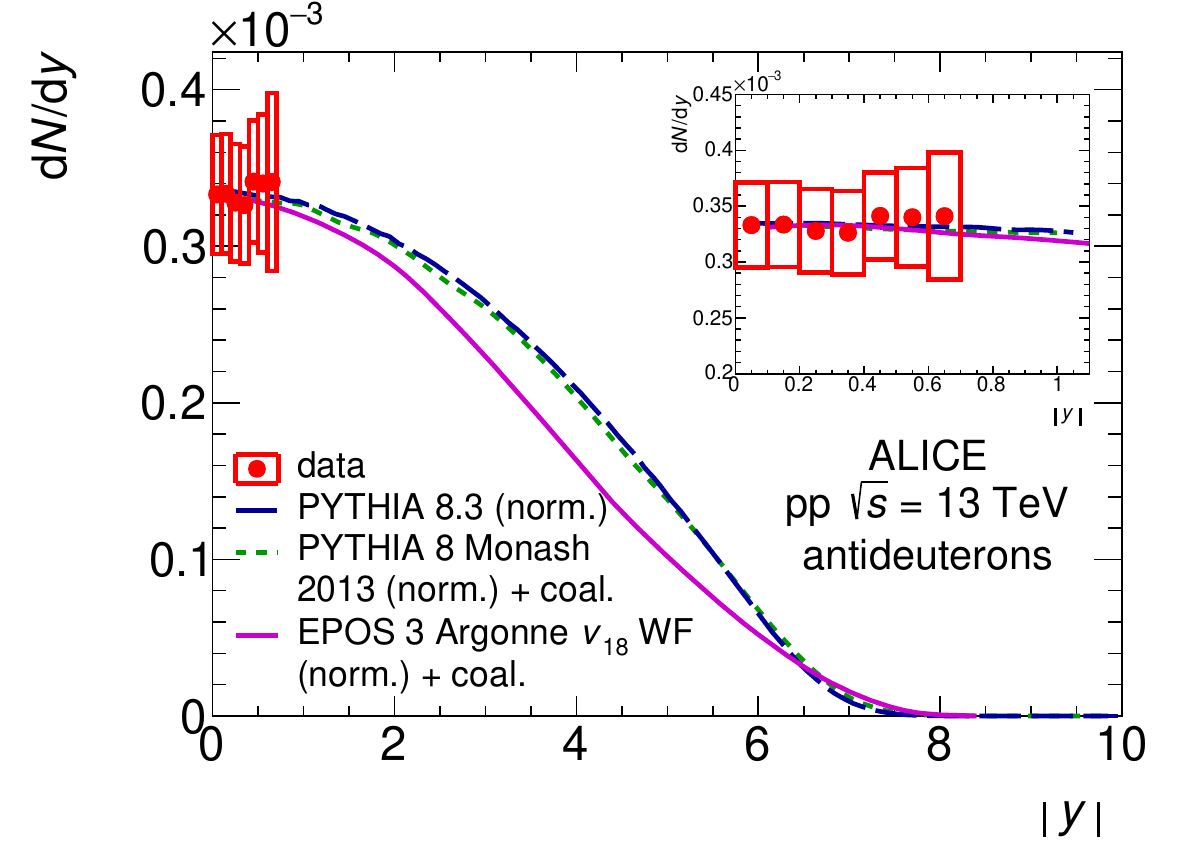}
\end{minipage}
\caption{Integrated yields of antiprotons (left) and antideuterons (right) as a function of rapidity, compared with the corresponding predictions of three models (see text for details). Statistical and systematic uncertainties are represented by vertical bars and boxes, respectively. The statistical uncertainties are smaller than the size of the markers in the reported scale and, hence, not visible. In the insets of the figures, a zoom in the low-rapidity region is displayed. The integrated yields estimated by models are normalized to the measured ones, see text for details. }
\label{fig:yields}
\end{figure}

The \pt-differential yields per unit of rapidity (d$^2N/$d$p_{\rm T}$d$y$) of antiprotons and antideuterons in MB events and analyzed rapidity intervals are shown in Fig.~\ref{fig:pt_spectra}. The results are consistent with those measured in $\left| y \right|<0.5$~\cite{ALICE:2020nkc,Acharya:2020sfy}. The final transverse-momentum distributions are fitted using the L\'evy--Tsallis function~\cite{Tsallis:1987eu} to extrapolate the yields to the unmeasured regions. The L\'evy--Tsallis fits of the spectra in the high-rapidity intervals ($|y| \geq 0.5$) are constrained to the spectrum in the lowest rapidity interval ($|y|<0.1$), such that the normalization is always adjusted during the fit while the shape is constrained to the spectrum at low rapidity, which covers a broader \pt range. This procedure assumes that the shape of the \pt spectra is independent of rapidity, and this assumption was validated by the rather stable $\chi^2/$NDF values of the fits across the studied rapidity interval. In order to obtain the integrated yields, the integrals of the data points of the transverse momentum spectra in the measured regions are summed to the integrals of the fit functions at low (down to 0) and high (up to 10 GeV/$c$) \pt. 
The resulting \pt-integrated yields (d$N$/d$y$) of antiprotons and antideuterons are shown in Fig.~\ref{fig:yields}. The measured integrated yields of both species show a flat trend with rapidity, up to $|y|=$ 0.7. The antiproton integrated yields as a function of rapidity are compared with the predictions of two event generators, namely \textsc{Pythia 8}~\cite{PYTHIA8Monasch2013}, with two different tunes (described in the following), and \mbox{\textsc{Epos} 3}~\cite{Werner:2010aa, Werner:2013tya}. The measured antideuteron yields are compared with the predictions of three models, based on different assumptions and making use of different event generators. Two of these models use \textsc{Pythia 8}~\cite{PYTHIA8Monasch2013} as event generator, the third \mbox{\textsc{Epos} 3}~\cite{Werner:2010aa, Werner:2013tya}. 
While for antiprotons the event generators produce the desired particles directly, there are no antideuterons natively in the MC and their production is included as an afterburner with the coalescence model, on top of the events generated using the two event generators. 
All models shown in Fig.~\ref{fig:yields} are normalized to data, by dividing the yields obtained with the simulations by a factor corresponding to the ratio between the measured yield and the simulated one in the lowest rapidity interval ($|y|<0.1$), since the natural abundances of particles are not well reproduced by the event generators. 

In the first approach, already used in Ref.~\cite{ALICE:2022ugx} and described in detail in the corresponding Supplemental Material~\cite{supplmat}, the phase space distributions of antinucleons are generated with \textsc{Pythia} 8 with the Monash 2013 tune~\cite{PYTHIA8Monasch2013}, and their \pt spectra are re-weighted to match the measured antiproton \pt spectra. In this simple model, all spatial correlations are ignored and the antideuteron is formed if an antiproton and an antineutron have a momentum difference below a given coalescence momentum \mbox{$\Delta p<p_{0}$} in the antideuteron rest frame. The best estimate of the coalescence momentum for this model is \mbox{$p_0=(285 \pm 1$) MeV/$c$}~\cite{ALICE:2022ugx}. 
The second approach is a reaction-based model~\cite{Bierlich:2022pfr}, where antideuterons are generated by ordinary nuclear reactions between antinucleons produced in the collision with parameterized energy-dependent cross sections tuned on available experimental data~\cite{Dal:2015sha}, as implemented in the MC event generator \textsc{Pythia} 8.3.
Both, the simple coalescence and the reaction-based model, are based on the simplistic assumption that coalescence can happen only if antiprotons and antineutrons are close in momentum space, while all spatial correlations are neglected, and the nucleus wave function does not come into play. The third approach is a state-of-the-art coalescence model~\cite{Mahlein:2023fmx} based on the Wigner function formalism, applied as an afterburner to the nucleons generated with the \mbox{\textsc{Epos} 3} event generator. In such a model, the quantum-mechanical properties of the nucleus are taken into account through its Wigner function. Antideuterons are formed following a probability calculated on an event-by-event basis, by folding the spatial distribution of nucleons with the Wigner function of the final bound state. The coalescence model is very sensitive to the source size, i.e., the size of the particle-emitting source that can be measured using femtoscopy with two-particle correlations~\cite{ALICE:2020ibs}. Since the one implemented in \mbox{\textsc{Epos} 3} does not reproduce the measurements~\cite{Mahlein:2023fmx}, a parameterization of the transverse-mass ($m_\mathrm{T}$) dependence of the measured source size for high-multiplicity events (\mbox{$r_0 = $ 1.249 fm}~\cite{ALICE:2019buq}) scaled to the minimum bias value (\mbox{$r_0= $ 1.18 fm}~\cite{ALICE:2019Oct}) was used. It has to be noted that the average values of transverse mass, $\langle m_{\rm T}\rangle$, of the high multiplicity sample (corresponding to the interval of 0--0.17\% in multiplicity) and of the minimum bias one (full multiplicity 0--100\%) differ by $\sim10\%$. Nonetheless, the source size is assumed to be independent of rapidity. Note also that the integrated yields predicted by the models are scaled to the data, while the shapes of the \pt distributions are not modified. Therefore, the models are no longer sensitive to the magnitude of the source size used for the calculations, but only to its $m_\mathrm{T}$ trend.
Additionally, the predictions are obtained using the Argonne $v_{18}$ wavefunction which was shown to give the best predictions for the antideuteron momentum distribution~\cite{Mahlein:2023fmx}. In this way, one obtains predictions using a realistic coalescence model, which considers not only the momentum distributions of nucleons but also their spatial correlations, as well as the quantum-mechanical nature of the coalescence process. 
The integrated yields of antiprotons and antideuterons as a function of rapidity obtained with the event generators and with the three models previously described show a similar trend, rather flat at low rapidities ($\left|y\right|<$ 1.5) and decreasing starting from rapidity about 1.5, up to the kinematic limit of $\left|y\right|\sim$ 9.5, related to the collision system and energy. However, while the predictions of the models based on \textsc{Pythia 8} and simple coalescence are roughly equal in the full rapidity interval (0--9.5), a large difference, for rapidity larger than 1.5, is observed between these models and the third one in which the distributions of the antinucleons are obtained using \mbox{\textsc{Epos} 3} and a state-of-the-art model for deuteron formation via coalescence. 

\begin{figure}[!hbt]
\centering
\begin{minipage}[b]{0.49\textwidth}
    \includegraphics[width=\textwidth]{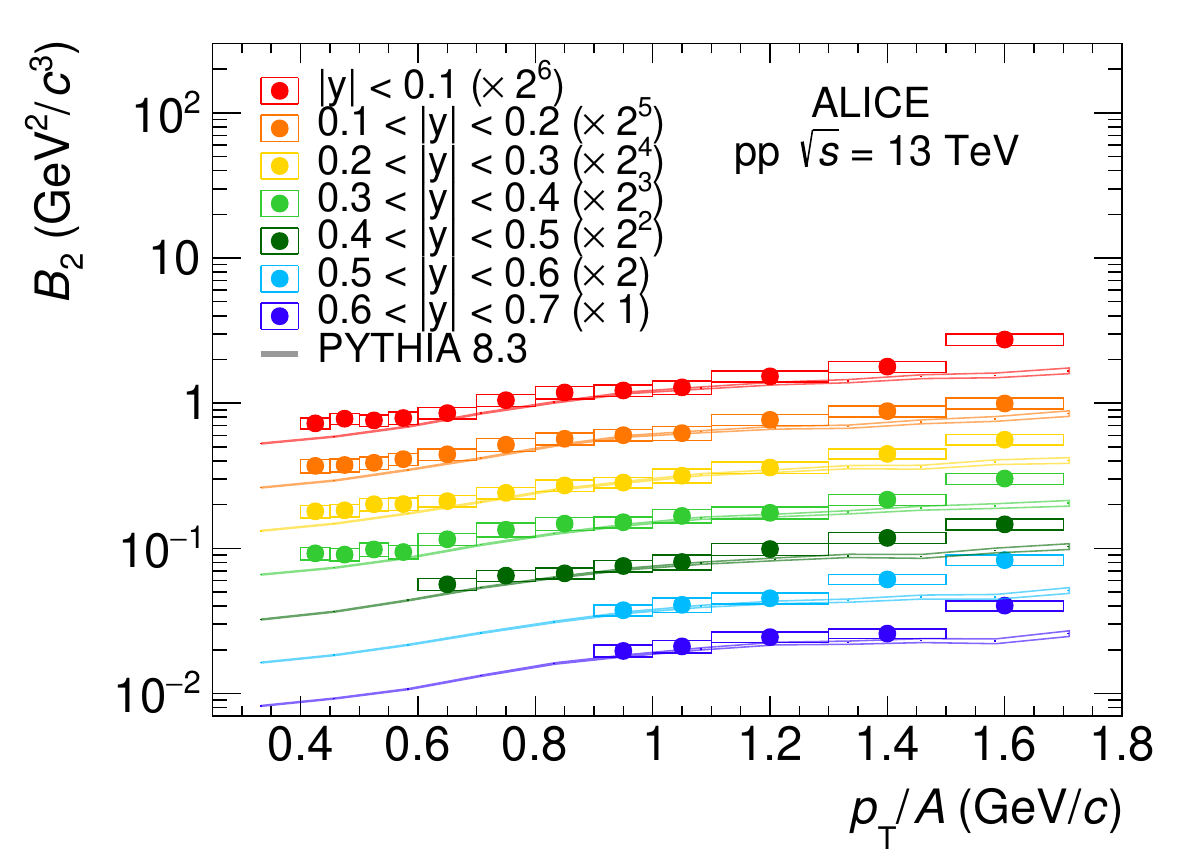}
\end{minipage}
  \hfill
\begin{minipage}[b]{0.49\textwidth}
	\includegraphics[width=\textwidth]{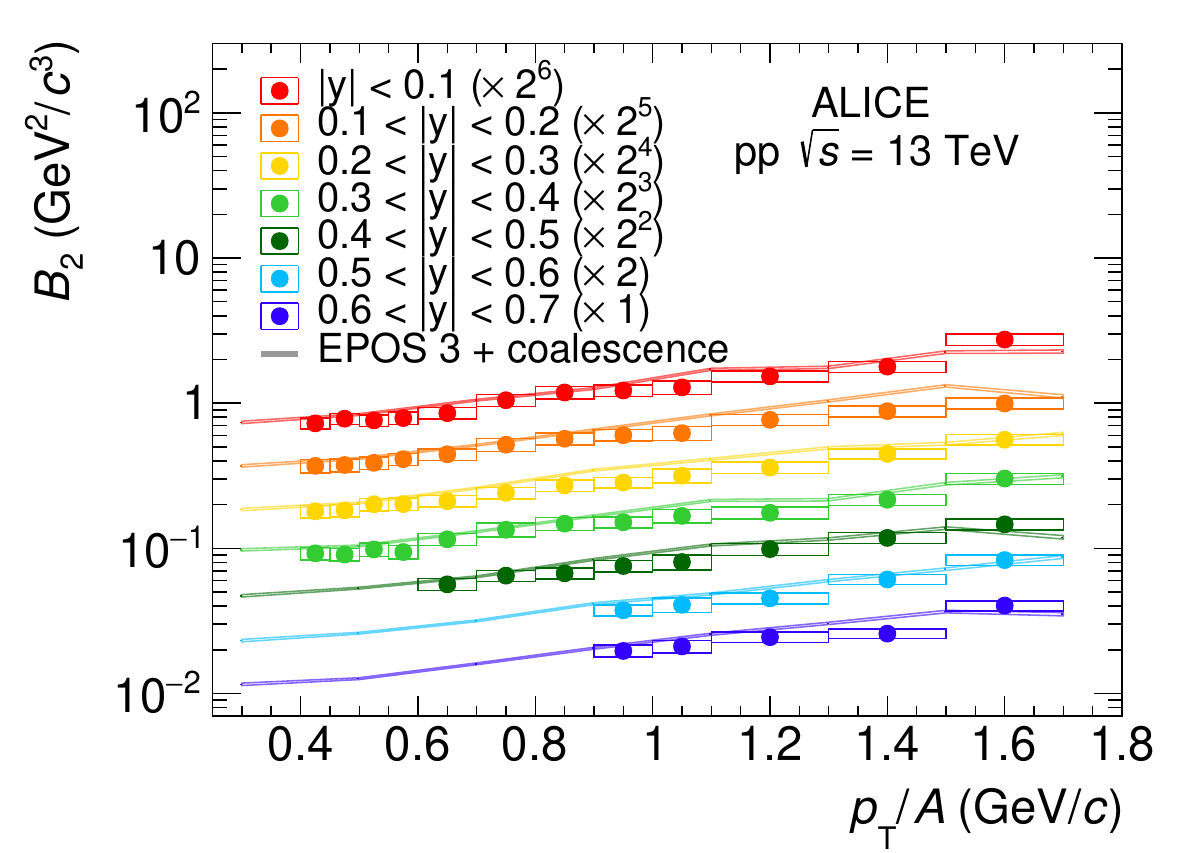}
\end{minipage}   
\caption{Coalescence parameter $B_2$ as a function of the transverse momentum per nucleon for different rapidity intervals. Data are compared to model predictions from \textsc{Pythia} 8.3 (left panel) and from a coalescence model~\cite{Mahlein:2023fmx} used as afterburner of \mbox{\textsc{Epos} 3} (right panel), shown as colored lines. Statistical and systematic uncertainties on the data points are represented by vertical bars and boxes, respectively. The statistical uncertainties of the data points are smaller than the size of the markers in the reported scale and, hence, not visible. }
\label{fig:B2}
\end{figure}

Using Eq.~\ref{eq:BA}, the coalescence parameters $B_2$ in the different rapidity classes are calculated and shown in Fig.~\ref{fig:B2}, in comparison with predictions from \textsc{Pythia} 8.3 (left panel) and from \mbox{\textsc{Epos} 3} with the coalescence model of Ref.~\cite{Mahlein:2023fmx} used as afterburner (right panel). For the computation of the uncertainties on the $B_2$ parameter, the invariant yields of antiprotons and antideuterons are considered uncorrelated and their uncertainties are propagated accordingly. The predictions obtained with \textsc{Pythia} 8 with the Monash 2013 tune have a negligible difference with respect to the predictions obtained with \textsc{Pythia} 8.3, and therefore not shown in the following. In the models, the antiproton \pt distributions are reweighted to the measured ones, while the antideuterons are only scaled to the integrated measured yields, and their \pt shapes are not changed.
The rising trend of the coalescence parameters with \pt/$A$ is well reproduced by the two models. Since larger coalescence parameters correspond to smaller source sizes (at higher \pt), this result reflects the observed decreasing trend of the source size with $\langle m_{\rm T}\rangle$~\cite{ALICE:2020ibs}. 

\begin{figure}[!hbt]
\centering
\begin{minipage}[b]{0.49\textwidth}
        \includegraphics[width=\textwidth]{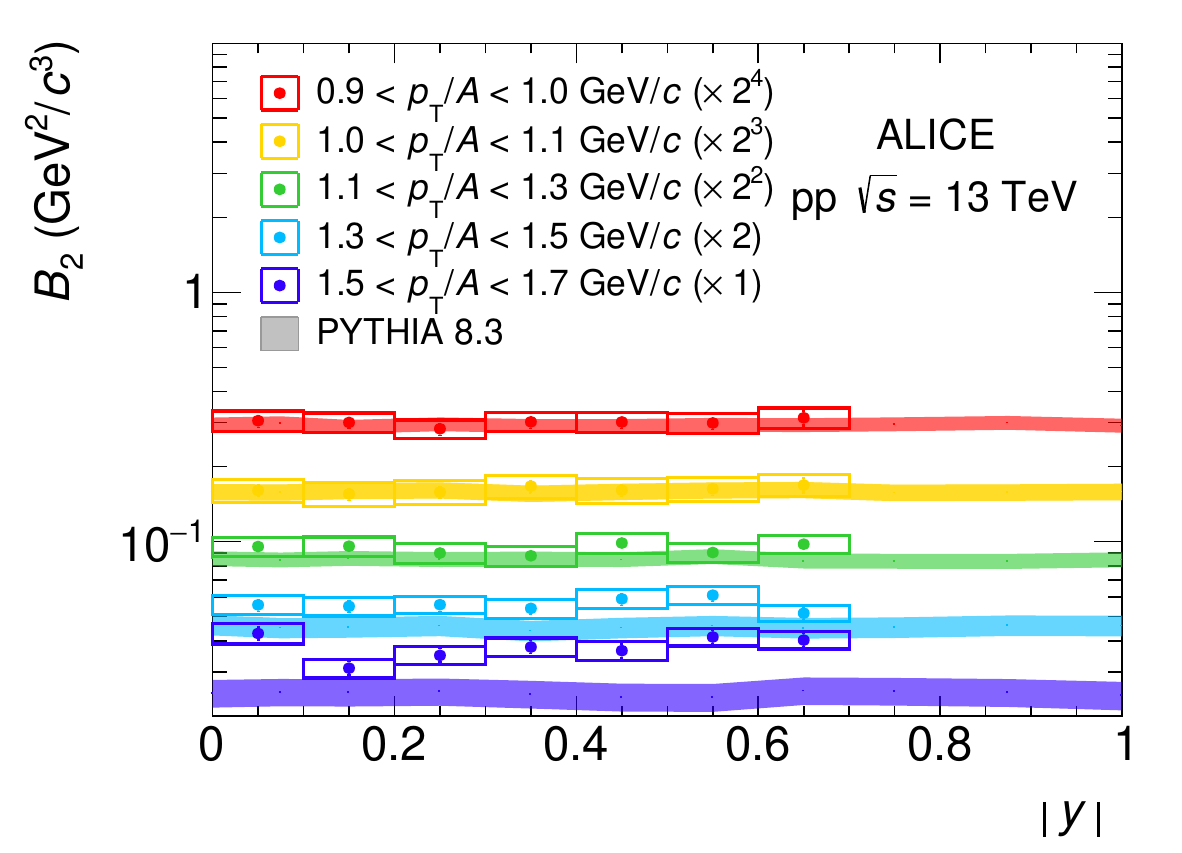}
 \end{minipage}
  \hfill
\begin{minipage}[b]{0.49\textwidth}
	\includegraphics[width=\textwidth]{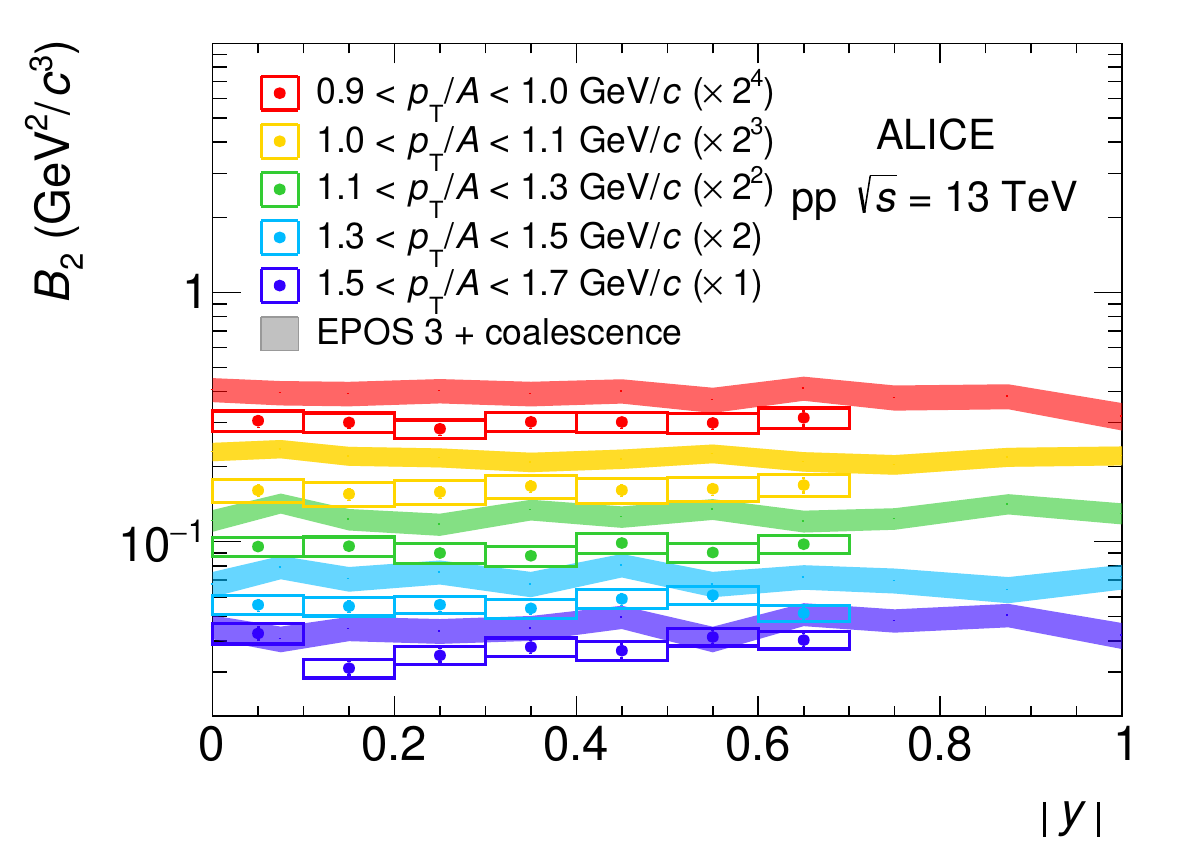}
\end{minipage}
\caption{Coalescence parameter $B_2$ as a function of rapidity, for fixed transverse momentum per nucleon values. Statistical and systematic uncertainties on the data points are represented by vertical bars and boxes, respectively. The statistical uncertainties are smaller than the size of the markers in the reported scale and, hence, not visible. Bands show the model predictions from \mbox{\textsc{Pythia} 8.3} (left panel) and from a coalescence model~\cite{Mahlein:2023fmx} used as afterburner of \mbox{\textsc{Epos} 3} (right panel), the width reflecting the statistical uncertainties. }
\label{fig:B2_fixedpT}
\end{figure}

Finally, selecting intervals of \pt/$A$, it is interesting to look at the trend of the coalescence parameter as a function of rapidity. The rapidity dependence of $B_2$ is shown in Fig.~\ref{fig:B2_fixedpT} and compared with predictions from \textsc{Pythia} 8.3 (left panel) and from a coalescence model~\cite{Mahlein:2023fmx} used as afterburner of \mbox{\textsc{Epos} 3} (right panel). As for the yields, the coalescence parameter is flat in the rapidity region investigated by the measurements. The models reproduce the measured trend, mostly with an agreement within 2$\sigma$. Some deviations beyond the 2$\sigma$ level are present at high \pt/$A$ ($>1.5$ GeV/$c$) for the model based on \mbox{\textsc{Pythia} 8.3}. 
A similar trend of $B_2$, flat as a function of rapidity, was also observed by the BRAHMS Collaboration in central Au--Au collisions at $\sqrt{s_{\rm NN}}=200$ GeV~\cite{Arsene:2010px}. 

The models shown in Figs.~\ref{fig:B2} and~\ref{fig:B2_fixedpT} have similar statistical uncertainties, of about 10\%, which are reflected in the width of the colored bands. 

The flux of antideuterons from cosmic rays as a function of rigidity, reported in Fig.~1 of Ref.~\cite{Blum:2023lmr}, is predicted with the model of Ref.~\cite{Blum:2017qnn} using the coalescence parameter $B_2$ measured at midrapidity in pp collisions at \mbox{$\sqrt{s}=$ 7 TeV} with ALICE~\cite{deuteron_pp_7TeV}. 
The rapidity and \pt dependencies of the coalescence parameters predicted by \textsc{Pythia} 8.3 and \mbox{\textsc{Epos} 3}, scaled to match the experimental measurements in the first \pt interval reported here, are extrapolated at high \pt (\pt $>$ 1.7 GeV/$c$) and forward rapidity ($|y|>$ 0.7) and provided to the authors of Refs.~\cite{Blum:2017qnn, Blum:2023lmr}. In these models, the $B_2$ distributions are substantially flat as a function of rapidity and increase monotonically as a function of \pt$/A$.
No difference is found in the shape of the antideuteron flux shown in Fig.~1 of Ref.~\cite{Blum:2023lmr}, either using the different models presented in this Letter or using the $B_2$ measured at midrapidity in pp collisions at \mbox{$\sqrt{s}=$ 7 TeV}~\cite{deuteron_pp_7TeV}. 
A constant $B_2$ as a function of rapidity is a common assumption in several CR models. However, in the coalescence models presented in this work, the trend with rapidity of $B_2$ is investigated for the first time. 
Additionally, in Ref.~\cite{Blum:2023lmr}, the author investigates the role of the rapidity extrapolation in predicting the cosmic ray flux of antinuclei (i.e., antiprotons, antideuterons, and antihelium-3), by setting the production cross section to zero in different kinematical regions. The results for the antideuteron flux from CR, presented in Fig.~1 of Ref.~\cite{Blum:2023lmr}, show that at low kinetic energies, corresponding to rigidity \mbox{$R\leq 5$ GV} (being the rigidity defined as $R= \frac{p c}{Z e}$, where $p$ is the momentum, $c$ is the speed of light, and $Ze$ is the charge of the cosmic-ray particle), about 90\% of the flux is due to antideuterons with rapidity larger than 0.5. This kinematical region is also the one where the CR flux is expected to be dominated by the DM signal, according to many cosmological models~\cite{Serksnyte}, and hence of high impact for astrophysics. Therefore, it is important to measure the coalescence parameter in the rapidity region $|y|>0.5$, in order to use experimental measurements as input for the astrophysical models. Moreover, the results of the antideuteron flux show that LHC experiments (such as LHCb~\cite{LHCb:2008vvz,LHCb:2014set,LHCb:2023tlg} in its fixed-target configuration, and the future ALICE 3 facility~\cite{ALICE:2022wwr}) are expected to fill the rapidity gap \mbox{(0.5 $<\left|y\right| <$ 1.5)} providing measurements of production yields of antinuclei. These measurements would be of important use to the astrophysics community for the calculation of the flux of antinuclei from cosmic rays, to avoid relying on the model extrapolation at forward rapidities.

\section{Summary}
\label{sec:Summary}
In this Letter, the \pt-differential yields of antideuterons and antiprotons as a function of rapidity are measured in pp collisions at \mbox{$\sqrt{s} =$ 13 TeV}, for the first time rapidity-differentially up to \mbox{$|y| =$ 0.7}. Using the measured spectra, the \pt and rapidity dependence of the coalescence parameter $B_2$ is explored. Both the antiproton and antideuteron integrated yields, and the $B_2$ are found to be independent of rapidity in the measured range. The rapidity dependence of the obtained yields and $B_2$ is extrapolated for $|y|>0.7$ using phenomenological antideuteron production models based on coalescence, making use of \textsc{Pythia} and \textsc{Epos} 3 as event generators. The resulting predictions of $B_2$ as a function of \pt and rapidity are used as input for the model of Ref.~\cite{Blum:2017qnn}, to investigate the role of the rapidity dependence of the coalescence parameter in the predictions of the flux of antinuclei from cosmic rays~\cite{Blum:2023lmr}. No impact on the predictions of the flux of antideuterons from CRs is seen when including the flat rapidity dependence of $B_2$ predicted with the models presented in this Letter, with respect to the predictions that use the $B_2$ measured at midrapidity in pp collisions at \mbox{$\sqrt{s}=$ 7 TeV}~\cite{deuteron_pp_7TeV}. However, as pointed out in Ref.~\cite{Blum:2023lmr}, at low kinetic energies ($E_{\rm{kin}}<$1 GeV/nucleon) about 90\% of the flux is due to antideuterons with rapidity between 0.5 and 1.5. Hence, the rapidity-dependent measurements of the production yields and coalescence probability of antinuclei in the region \mbox{0.5 $<\left|y\right| <$ 1.5} are a crucial input for CR models. These measurements have a fundamental impact on astrophysical indirect searches for dark matter, as the antinuclei produced from CR interactions are the dominant background source in the region of low kinetic energies. 
This measurement can be extended in the rapidity interval of interest by other facilities, present or future, such as LHCb~\cite{LHCb:2008vvz,LHCb:2023tlg} and ALICE 3~\cite{ALICE:2022wwr}. 


\newenvironment{acknowledgement}{\relax}{\relax}
\begin{acknowledgement}
\section*{Acknowledgements}

The ALICE Collaboration would like to thank all its engineers and technicians for their invaluable contributions to the construction of the experiment and the CERN accelerator teams for the outstanding performance of the LHC complex.
The ALICE Collaboration gratefully acknowledges the resources and support provided by all Grid centres and the Worldwide LHC Computing Grid (WLCG) collaboration.
The ALICE Collaboration acknowledges the following funding agencies for their support in building and running the ALICE detector:
A. I. Alikhanyan National Science Laboratory (Yerevan Physics Institute) Foundation (ANSL), State Committee of Science and World Federation of Scientists (WFS), Armenia;
Austrian Academy of Sciences, Austrian Science Fund (FWF): [M 2467-N36] and Nationalstiftung f\"{u}r Forschung, Technologie und Entwicklung, Austria;
Ministry of Communications and High Technologies, National Nuclear Research Center, Azerbaijan;
Conselho Nacional de Desenvolvimento Cient\'{\i}fico e Tecnol\'{o}gico (CNPq), Financiadora de Estudos e Projetos (Finep), Funda\c{c}\~{a}o de Amparo \`{a} Pesquisa do Estado de S\~{a}o Paulo (FAPESP) and Universidade Federal do Rio Grande do Sul (UFRGS), Brazil;
Bulgarian Ministry of Education and Science, within the National Roadmap for Research Infrastructures 2020-2027 (object CERN), Bulgaria;
Ministry of Education of China (MOEC) , Ministry of Science \& Technology of China (MSTC) and National Natural Science Foundation of China (NSFC), China;
Ministry of Science and Education and Croatian Science Foundation, Croatia;
Centro de Aplicaciones Tecnol\'{o}gicas y Desarrollo Nuclear (CEADEN), Cubaenerg\'{\i}a, Cuba;
Ministry of Education, Youth and Sports of the Czech Republic, Czech Republic;
The Danish Council for Independent Research | Natural Sciences, the VILLUM FONDEN and Danish National Research Foundation (DNRF), Denmark;
Helsinki Institute of Physics (HIP), Finland;
Commissariat \`{a} l'Energie Atomique (CEA) and Institut National de Physique Nucl\'{e}aire et de Physique des Particules (IN2P3) and Centre National de la Recherche Scientifique (CNRS), France;
Bundesministerium f\"{u}r Bildung und Forschung (BMBF) and GSI Helmholtzzentrum f\"{u}r Schwerionenforschung GmbH, Germany;
General Secretariat for Research and Technology, Ministry of Education, Research and Religions, Greece;
National Research, Development and Innovation Office, Hungary;
Department of Atomic Energy Government of India (DAE), Department of Science and Technology, Government of India (DST), University Grants Commission, Government of India (UGC) and Council of Scientific and Industrial Research (CSIR), India;
National Research and Innovation Agency - BRIN, Indonesia;
Istituto Nazionale di Fisica Nucleare (INFN), Italy;
Japanese Ministry of Education, Culture, Sports, Science and Technology (MEXT) and Japan Society for the Promotion of Science (JSPS) KAKENHI, Japan;
Consejo Nacional de Ciencia (CONACYT) y Tecnolog\'{i}a, through Fondo de Cooperaci\'{o}n Internacional en Ciencia y Tecnolog\'{i}a (FONCICYT) and Direcci\'{o}n General de Asuntos del Personal Academico (DGAPA), Mexico;
Nederlandse Organisatie voor Wetenschappelijk Onderzoek (NWO), Netherlands;
The Research Council of Norway, Norway;
Pontificia Universidad Cat\'{o}lica del Per\'{u}, Peru;
Ministry of Science and Higher Education, National Science Centre and WUT ID-UB, Poland;
Korea Institute of Science and Technology Information and National Research Foundation of Korea (NRF), Republic of Korea;
Ministry of Education and Scientific Research, Institute of Atomic Physics, Ministry of Research and Innovation and Institute of Atomic Physics and Universitatea Nationala de Stiinta si Tehnologie Politehnica Bucuresti, Romania;
Ministry of Education, Science, Research and Sport of the Slovak Republic, Slovakia;
National Research Foundation of South Africa, South Africa;
Swedish Research Council (VR) and Knut \& Alice Wallenberg Foundation (KAW), Sweden;
European Organization for Nuclear Research, Switzerland;
Suranaree University of Technology (SUT), National Science and Technology Development Agency (NSTDA) and National Science, Research and Innovation Fund (NSRF via PMU-B B05F650021), Thailand;
Turkish Energy, Nuclear and Mineral Research Agency (TENMAK), Turkey;
National Academy of  Sciences of Ukraine, Ukraine;
Science and Technology Facilities Council (STFC), United Kingdom;
National Science Foundation of the United States of America (NSF) and United States Department of Energy, Office of Nuclear Physics (DOE NP), United States of America.
In addition, individual groups or members have received support from:
Czech Science Foundation (grant no. 23-07499S), Czech Republic;
European Research Council (grant no. 950692), European Union;
ICSC - Centro Nazionale di Ricerca in High Performance Computing, Big Data and Quantum Computing, European Union - NextGenerationEU;
Academy of Finland (Center of Excellence in Quark Matter) (grant nos. 346327, 346328), Finland.

\end{acknowledgement}

\bibliographystyle{utphys}   
\bibliography{bibliography}

\newpage
\appendix

\section{The ALICE Collaboration}
\label{app:collab}
\begin{flushleft} 
\small

S.~Acharya\,\orcidlink{0000-0002-9213-5329}\,$^{\rm 127}$, 
D.~Adamov\'{a}\,\orcidlink{0000-0002-0504-7428}\,$^{\rm 86}$, 
A.~Agarwal$^{\rm 135}$, 
G.~Aglieri Rinella\,\orcidlink{0000-0002-9611-3696}\,$^{\rm 32}$, 
L.~Aglietta\,\orcidlink{0009-0003-0763-6802}\,$^{\rm 24}$, 
M.~Agnello\,\orcidlink{0000-0002-0760-5075}\,$^{\rm 29}$, 
N.~Agrawal\,\orcidlink{0000-0003-0348-9836}\,$^{\rm 25}$, 
Z.~Ahammed\,\orcidlink{0000-0001-5241-7412}\,$^{\rm 135}$, 
S.~Ahmad\,\orcidlink{0000-0003-0497-5705}\,$^{\rm 15}$, 
S.U.~Ahn\,\orcidlink{0000-0001-8847-489X}\,$^{\rm 71}$
% ...

\section*{Affiliation Notes}

$^{\rm I}$ Deceased\\
$^{\rm II}$ Also at: Max-Planck-Institut fur Physik, Munich, Germany\\
$^{\rm III}$ Also at: Italian National Agency for New Technologies, Energy and Sustainable Economic Development (ENEA), Bologna, Italy\\
$^{\rm IV}$ Also at: Dipartimento DET del Politecnico di Torino, Turin, Italy\\
$^{\rm V}$ Also at: Department of Applied Physics, Aligarh Muslim University, Aligarh, India\\
$^{\rm VI}$ Also at: Institute of Theoretical Physics, University of Wroclaw, Poland\\
$^{\rm VII}$ Also at: Facultad de Ciencias, Universidad Nacional Autónoma de México, Mexico City, Mexico\\
$^{\rm VIII}$ Also at: An institution covered by a cooperation agreement with CERN\\

\section*{Collaboration Institutes}

$^{1}$ A.I. Alikhanyan National Science Laboratory (Yerevan Physics Institute) Foundation, Yerevan, Armenia\\
$^{2}$ AGH University of Krakow, Cracow, Poland\\
$^{3}$ Bogolyubov Institute for Theoretical Physics, National Academy of Sciences of Ukraine, Kiev, Ukraine\\
$^{4}$ Bose Institute, Department of Physics  and Centre for Astroparticle Physics and Space Science (CAPSS), Kolkata, India\\
$^{5}$ California Polytechnic State University, San Luis Obispo, California, United States\\
$^{6}$ Central China Normal University, Wuhan, China\\
$^{7}$ Centro de Aplicaciones Tecnol\'{o}gicas y Desarrollo Nuclear (CEADEN), Havana, Cuba\\
$^{8}$ Centro de Investigaci\'{o}n y de Estudios Avanzados (CINVESTAV), Mexico City and M\'{e}rida, Mexico\\
$^{9}$ Chicago State University, Chicago, Illinois, United States\\
$^{10}$ China Institute of Atomic Energy, Beijing, China\\
$^{11}$ China University of Geosciences, Wuhan, China\\
$^{12}$ Chungbuk National University, Cheongju, Republic of Korea\\
$^{13}$ Comenius University Bratislava, Faculty of Mathematics, Physics and Informatics, Bratislava, Slovak Republic\\
$^{14}$ Creighton University, Omaha, Nebraska, United States\\
$^{15}$ Department of Physics, Aligarh Muslim University, Aligarh, India\\
$^{16}$ Department of Physics, Pusan National University, Pusan, Republic of Korea\\
$^{17}$ Department of Physics, Sejong University, Seoul, Republic of Korea\\
$^{18}$ Department of Physics, University of California, Berkeley, California, United States\\
$^{19}$ Department of Physics, University of Oslo, Oslo, Norway\\
$^{20}$ Department of Physics and Technology, University of Bergen, Bergen, Norway\\
$^{21}$ Dipartimento di Fisica, Universit\`{a} di Pavia, Pavia, Italy\\
$^{22}$ Dipartimento di Fisica dell'Universit\`{a} and Sezione INFN, Cagliari, Italy\\
$^{23}$ Dipartimento di Fisica dell'Universit\`{a} and Sezione INFN, Trieste, Italy\\
$^{24}$ Dipartimento di Fisica dell'Universit\`{a} and Sezione INFN, Turin, Italy\\
$^{25}$ Dipartimento di Fisica e Astronomia dell'Universit\`{a} and Sezione INFN, Bologna, Italy\\
$^{26}$ Dipartimento di Fisica e Astronomia dell'Universit\`{a} and Sezione INFN, Catania, Italy\\
$^{27}$ Dipartimento di Fisica e Astronomia dell'Universit\`{a} and Sezione INFN, Padova, Italy\\
$^{28}$ Dipartimento di Fisica `E.R.~Caianiello' dell'Universit\`{a} and Gruppo Collegato INFN, Salerno, Italy\\
$^{29}$ Dipartimento DISAT del Politecnico and Sezione INFN, Turin, Italy\\
$^{30}$ Dipartimento di Scienze MIFT, Universit\`{a} di Messina, Messina, Italy\\
$^{31}$ Dipartimento Interateneo di Fisica `M.~Merlin' and Sezione INFN, Bari, Italy\\
$^{32}$ European Organization for Nuclear Research (CERN), Geneva, Switzerland\\
$^{33}$ Faculty of Electrical Engineering, Mechanical Engineering and Naval Architecture, University of Split, Split, Croatia\\
$^{34}$ Faculty of Nuclear Sciences and Physical Engineering, Czech Technical University in Prague, Prague, Czech Republic\\
$^{35}$ Faculty of Physics, Sofia University, Sofia, Bulgaria\\
$^{36}$ Faculty of Science, P.J.~\v{S}af\'{a}rik University, Ko\v{s}ice, Slovak Republic\\
$^{37}$ Faculty of Technology, Environmental and Social Sciences, Bergen, Norway\\
$^{38}$ Frankfurt Institute for Advanced Studies, Johann Wolfgang Goethe-Universit\"{a}t Frankfurt, Frankfurt, Germany\\
$^{39}$ Fudan University, Shanghai, China\\
$^{40}$ Gangneung-Wonju National University, Gangneung, Republic of Korea\\
$^{41}$ Gauhati University, Department of Physics, Guwahati, India\\
$^{42}$ Helmholtz-Institut f\"{u}r Strahlen- und Kernphysik, Rheinische Friedrich-Wilhelms-Universit\"{a}t Bonn, Bonn, Germany\\
$^{43}$ Helsinki Institute of Physics (HIP), Helsinki, Finland\\
$^{44}$ High Energy Physics Group,  Universidad Aut\'{o}noma de Puebla, Puebla, Mexico\\
$^{45}$ Horia Hulubei National Institute of Physics and Nuclear Engineering, Bucharest, Romania\\
$^{46}$ HUN-REN Wigner Research Centre for Physics, Budapest, Hungary\\
$^{47}$ Indian Institute of Technology Bombay (IIT), Mumbai, India\\
$^{48}$ Indian Institute of Technology Indore, Indore, India\\
$^{49}$ INFN, Laboratori Nazionali di Frascati, Frascati, Italy\\
$^{50}$ INFN, Sezione di Bari, Bari, Italy\\
$^{51}$ INFN, Sezione di Bologna, Bologna, Italy\\
$^{52}$ INFN, Sezione di Cagliari, Cagliari, Italy\\
$^{53}$ INFN, Sezione di Catania, Catania, Italy\\
$^{54}$ INFN, Sezione di Padova, Padova, Italy\\
$^{55}$ INFN, Sezione di Pavia, Pavia, Italy\\
$^{56}$ INFN, Sezione di Torino, Turin, Italy\\
$^{57}$ INFN, Sezione di Trieste, Trieste, Italy\\
$^{58}$ Inha University, Incheon, Republic of Korea\\
$^{59}$ Institute for Gravitational and Subatomic Physics (GRASP), Utrecht University/Nikhef, Utrecht, Netherlands\\
$^{60}$ Institute of Experimental Physics, Slovak Academy of Sciences, Ko\v{s}ice, Slovak Republic\\
$^{61}$ Institute of Physics, Homi Bhabha National Institute, Bhubaneswar, India\\
$^{62}$ Institute of Physics of the Czech Academy of Sciences, Prague, Czech Republic\\
$^{63}$ Institute of Space Science (ISS), Bucharest, Romania\\
$^{64}$ Institut f\"{u}r Kernphysik, Johann Wolfgang Goethe-Universit\"{a}t Frankfurt, Frankfurt, Germany\\
$^{65}$ Instituto de Ciencias Nucleares, Universidad Nacional Aut\'{o}noma de M\'{e}xico, Mexico City, Mexico\\
$^{66}$ Instituto de F\'{i}sica, Universidade Federal do Rio Grande do Sul (UFRGS), Porto Alegre, Brazil\\
$^{67}$ Instituto de F\'{\i}sica, Universidad Nacional Aut\'{o}noma de M\'{e}xico, Mexico City, Mexico\\
$^{68}$ iThemba LABS, National Research Foundation, Somerset West, South Africa\\
$^{69}$ Jeonbuk National University, Jeonju, Republic of Korea\\
$^{70}$ Johann-Wolfgang-Goethe Universit\"{a}t Frankfurt Institut f\"{u}r Informatik, Fachbereich Informatik und Mathematik, Frankfurt, Germany\\
$^{71}$ Korea Institute of Science and Technology Information, Daejeon, Republic of Korea\\
$^{72}$ KTO Karatay University, Konya, Turkey\\
$^{73}$ Laboratoire de Physique Subatomique et de Cosmologie, Universit\'{e} Grenoble-Alpes, CNRS-IN2P3, Grenoble, France\\
$^{74}$ Lawrence Berkeley National Laboratory, Berkeley, California, United States\\
$^{75}$ Lund University Department of Physics, Division of Particle Physics, Lund, Sweden\\
$^{76}$ Nagasaki Institute of Applied Science, Nagasaki, Japan\\
$^{77}$ Nara Women{'}s University (NWU), Nara, Japan\\
$^{78}$ National and Kapodistrian University of Athens, School of Science, Department of Physics , Athens, Greece\\
$^{79}$ National Centre for Nuclear Research, Warsaw, Poland\\
$^{80}$ National Institute of Science Education and Research, Homi Bhabha National Institute, Jatni, India\\
$^{81}$ National Nuclear Research Center, Baku, Azerbaijan\\
$^{82}$ National Research and Innovation Agency - BRIN, Jakarta, Indonesia\\
$^{83}$ Niels Bohr Institute, University of Copenhagen, Copenhagen, Denmark\\
$^{84}$ Nikhef, National institute for subatomic physics, Amsterdam, Netherlands\\
$^{85}$ Nuclear Physics Group, STFC Daresbury Laboratory, Daresbury, United Kingdom\\
$^{86}$ Nuclear Physics Institute of the Czech Academy of Sciences, Husinec-\v{R}e\v{z}, Czech Republic\\
$^{87}$ Oak Ridge National Laboratory, Oak Ridge, Tennessee, United States\\
$^{88}$ Ohio State University, Columbus, Ohio, United States\\
$^{89}$ Physics department, Faculty of science, University of Zagreb, Zagreb, Croatia\\
$^{90}$ Physics Department, Panjab University, Chandigarh, India\\
$^{91}$ Physics Department, University of Jammu, Jammu, India\\
$^{92}$ Physics Program and International Institute for Sustainability with Knotted Chiral Meta Matter (WPI-SKCM$^{2}$), Hiroshima University, Hiroshima, Japan\\
$^{93}$ Physikalisches Institut, Eberhard-Karls-Universit\"{a}t T\"{u}bingen, T\"{u}bingen, Germany\\
$^{94}$ Physikalisches Institut, Ruprecht-Karls-Universit\"{a}t Heidelberg, Heidelberg, Germany\\
$^{95}$ Physik Department, Technische Universit\"{a}t M\"{u}nchen, Munich, Germany\\
$^{96}$ Politecnico di Bari and Sezione INFN, Bari, Italy\\
$^{97}$ Research Division and ExtreMe Matter Institute EMMI, GSI Helmholtzzentrum f\"ur Schwerionenforschung GmbH, Darmstadt, Germany\\
$^{98}$ Saga University, Saga, Japan\\
$^{99}$ Saha Institute of Nuclear Physics, Homi Bhabha National Institute, Kolkata, India\\
$^{100}$ School of Physics and Astronomy, University of Birmingham, Birmingham, United Kingdom\\
$^{101}$ Secci\'{o}n F\'{\i}sica, Departamento de Ciencias, Pontificia Universidad Cat\'{o}lica del Per\'{u}, Lima, Peru\\
$^{102}$ Stefan Meyer Institut f\"{u}r Subatomare Physik (SMI), Vienna, Austria\\
$^{103}$ SUBATECH, IMT Atlantique, Nantes Universit\'{e}, CNRS-IN2P3, Nantes, France\\
$^{104}$ Sungkyunkwan University, Suwon City, Republic of Korea\\
$^{105}$ Suranaree University of Technology, Nakhon Ratchasima, Thailand\\
$^{106}$ Technical University of Ko\v{s}ice, Ko\v{s}ice, Slovak Republic\\
$^{107}$ The Henryk Niewodniczanski Institute of Nuclear Physics, Polish Academy of Sciences, Cracow, Poland\\
$^{108}$ The University of Texas at Austin, Austin, Texas, United States\\
$^{109}$ Universidad Aut\'{o}noma de Sinaloa, Culiac\'{a}n, Mexico\\
$^{110}$ Universidade de S\~{a}o Paulo (USP), S\~{a}o Paulo, Brazil\\
$^{111}$ Universidade Estadual de Campinas (UNICAMP), Campinas, Brazil\\
$^{112}$ Universidade Federal do ABC, Santo Andre, Brazil\\
$^{113}$ Universitatea Nationala de Stiinta si Tehnologie Politehnica Bucuresti, Bucharest, Romania\\
$^{114}$ University of Cape Town, Cape Town, South Africa\\
$^{115}$ University of Derby, Derby, United Kingdom\\
$^{116}$ University of Houston, Houston, Texas, United States\\
$^{117}$ University of Jyv\"{a}skyl\"{a}, Jyv\"{a}skyl\"{a}, Finland\\
$^{118}$ University of Kansas, Lawrence, Kansas, United States\\
$^{119}$ University of Liverpool, Liverpool, United Kingdom\\
$^{120}$ University of Science and Technology of China, Hefei, China\\
$^{121}$ University of South-Eastern Norway, Kongsberg, Norway\\
$^{122}$ University of Tennessee, Knoxville, Tennessee, United States\\
$^{123}$ University of the Witwatersrand, Johannesburg, South Africa\\
$^{124}$ University of Tokyo, Tokyo, Japan\\
$^{125}$ University of Tsukuba, Tsukuba, Japan\\
$^{126}$ Universit\"{a}t M\"{u}nster, Institut f\"{u}r Kernphysik, M\"{u}nster, Germany\\
$^{127}$ Universit\'{e} Clermont Auvergne, CNRS/IN2P3, LPC, Clermont-Ferrand, France\\
$^{128}$ Universit\'{e} de Lyon, CNRS/IN2P3, Institut de Physique des 2 Infinis de Lyon, Lyon, France\\
$^{129}$ Universit\'{e} de Strasbourg, CNRS, IPHC UMR 7178, F-67000 Strasbourg, France, Strasbourg, France\\
$^{130}$ Universit\'{e} Paris-Saclay, Centre d'Etudes de Saclay (CEA), IRFU, D\'{e}partment de Physique Nucl\'{e}aire (DPhN), Saclay, France\\
$^{131}$ Universit\'{e}  Paris-Saclay, CNRS/IN2P3, IJCLab, Orsay, France\\
$^{132}$ Universit\`{a} degli Studi di Foggia, Foggia, Italy\\
$^{133}$ Universit\`{a} del Piemonte Orientale, Vercelli, Italy\\
$^{134}$ Universit\`{a} di Brescia, Brescia, Italy\\
$^{135}$ Variable Energy Cyclotron Centre, Homi Bhabha National Institute, Kolkata, India\\
$^{136}$ Warsaw University of Technology, Warsaw, Poland\\
$^{137}$ Wayne State University, Detroit, Michigan, United States\\
$^{138}$ Yale University, New Haven, Connecticut, United States\\
$^{139}$ Yonsei University, Seoul, Republic of Korea\\
$^{140}$  Zentrum  f\"{u}r Technologie und Transfer (ZTT), Worms, Germany\\
$^{141}$ Affiliated with an institute covered by a cooperation agreement with CERN\\
$^{142}$ Affiliated with an international laboratory covered by a cooperation agreement with CERN.\\

\end{flushleft} 

\end{document}